\documentclass[]{mn2e}
\usepackage{epsfig}

\title[Host of a peculiar AGN]{An evolutionary missing link? A modest-mass
early-type galaxy hosting an over-sized nuclear black hole}
\author[van Loon \& Sansom]{
      Jacco Th.\ van Loon$^{1}$ and Anne E.\ Sansom$^{2}$\\
$^{1}$Lennard-Jones Laboratories, Keele University, Staffordshire ST5 5BG,
      UK\\
$^{2}$Jeremiah Horrocks Institute, University of Central Lancashire, Preston
      PR1 2HE, UK}
\date{2015}
\pagerange{\pageref{firstpage}--\pageref{lastpage}}
\pubyear{2015}
\begin{document}
\maketitle
\label{firstpage}
\begin{abstract}
SAGE1C\,J053634.78$-$722658.5 is a galaxy at redshift $z=0.14$, discovered
behind the Large Magellanic Cloud in the {\it Spitzer} Space Telescope
``Surveying the Agents of Galaxy Evolution'' Spectroscopy survey (SAGE-Spec).
It has very strong silicate emission at 10 $\mu$m but negligible far-IR and UV
emission. This makes it a candidate for a bare AGN source in the IR, perhaps
seen pole-on, without significant IR emission from the host galaxy. In this
paper we present optical spectra taken with the Southern African Large
Telescope (SALT) to investigate the nature of the underlying host galaxy and
its AGN. We find broad H$\alpha$ emission characteristic of an AGN, plus
absorption lines associated with a mature stellar population ($>9$ Gyr), and
refine its redshift determination to $z=0.1428\pm0.0001$. There is no evidence
for any emission lines associated with star formation. This remarkable object
exemplifies the need for separating the emission from any AGN from that of the
host galaxy when employing infrared diagnostic diagrams. We estimate the black
hole mass, $M_{\rm BH}=3.5\pm0.8\times10^8$ M$_\odot$, host galaxy mass,
$M_{\rm stars}=2.5^{2.5}_{1.2}\times10^{10}$ M$_\odot$, and accretion luminosity,
$L_{\rm bol}({\rm AGN})=5.3\pm0.4\times10^{45}$ erg s$^{-1}$ ($\approx12$ per
cent of the Eddington luminosity) and find the AGN to be more prominent than
expected for a host galaxy of this modest size. The old age is in tension with
the downsizing paradigm in which this galaxy would recently have transformed
from a star-forming disc galaxy into an early-type, passively evolving galaxy.
\end{abstract}
\begin{keywords}
quasars: individual: SAGE1C\,J053634.78$-$722658.5 --
quasars: supermassive black holes --
galaxies: individual: [GC2009]\,J053642.29$-$722556.6 --
galaxies: peculiar --
galaxies: active --
stars: individual: [VS2015]\,J053640.54$-$722615.5
\end{keywords}

%=========================================================================== 1
\section{Introduction}

Active Galactic Nuclei (AGN) result from accretion of gas by a supermassive
black hole in the centre of a galaxy. AGN have been found in spiral galaxies
and massive elliptical galaxies, but whilst it may be assumed that all spiral
and massive elliptical galaxies harbour a supermassive black hole they do not
all exhibit the same level of nuclear activity. AGN activity is linked to the
assembly history of galaxies, with massive galaxies evolving faster. A result
is the intimate link between the mass of the black hole and that of the host
galaxy's spheroidal component (Kormendy \& Ho 2013; Graham 2015).

The current paradigm dictates that AGN manifest themselves in different guises
depending on the angle under which they are seen (Netzer 2015): the broad
emission-line region (BLR) in the dense, fast-rotating inner part of the AGN
is obscured from view by a dusty torus if seen under a high inclination; this
leaves exposed a narrow emission-line region (NLR). When the nucleus is seen
more directly, then the optical emission has a strong continuum component
obliterating the NLR -- these so-called ``blazars'' are relatively faint at
far-infrared (far-IR) wavelengths.

SAGE1C\,J053634.78$-$722658.5 (hereafter referred to as ``SAGE0536AGN'') is an
AGN which was serendipitously discovered by Hony et al.\ (2011, hereafter H11)
in the {\it Spitzer} Space Telescope Survey of the Agents of Galaxy Evolution
Spectroscopic followup of IR sources seen towards the Large Magellanic Cloud
(SAGE-Spec: Kemper et al.\ 2010; Woods et al.\ 2011). Its peculiarity, and
reason to have found its way into the SAGE-Spec target list, is due to its
spectral energy distribution (SED) peaking at mid-IR wavelengths; the
low-resolution {\it Spitzer} spectrum revealed the strongest silicate emission
of any known AGN (H11). It is faint at optical wavelengths, $\sim17$--18 mag,
and not detected at far-IR wavelengths, 70--160 $\mu$m, but radio loud (see
H11 for a summary of measurements). H11 used the weak mid-IR emission bands
that are generally attributed to Poly-cyclic Aromatic Hydro-carbons (PAHs) to
estimate a redshift $z=0.140\pm0.005$. They suggest the near-IR continuum
arises from hot dust near the accretion disc whilst the silicate emission
originates further out in dusty clouds within $\sim10$ pc from the black hole.
Silicate would appear in emission either if the torus is transparent and/or
clumpy, or the viewing angle is close to pole-on.

The unique character of this object merits further study, in particular aimed
at detecting the BLR and/or NLR and to determine the nature of its host
galaxy. We here present answers to those questions based on the first optical
spectroscopy of this intriguing AGN, which we obtained with the Southern
African Large Telescope (SALT).

%=========================================================================== 2
\section{Observations and data processing}

%
% TABLE 1
%
\begin{table*}
\caption{Observing parameters.}
\begin{tabular}{ccccccccc}
\hline\hline
dd/mm/yy                                       &
grating/filter                                 &
slit                                           &
$R=\lambda/\delta\lambda$                      &
\llap{g}rating/camera angl\rlap{e}             &
$\Delta\lambda$ (\AA)                          &
$t_{\rm int}$ (min)                             &
airmass                                        &
seeing                                         \\
\hline
08/09/12                                       &
PG0300/PC03850                                 &
$1\rlap{.}^{\prime\prime}25\times8^\prime$        &
300--700                                       &
$5\rlap{.}^\circ38$/$10\rlap{.}^\circ74$        &
3800--10400                                    &
$1\times$ 9                                    &
2.12                                           &
$1\rlap{.}^{\prime\prime}2$                      \\
11/10/12                                       &
PG1800/PC04600                                 &
$1\rlap{.}^{\prime\prime}25\times8^\prime$        &
3400--4000                                     &
\llap{4}$0\rlap{.}^\circ25$/$80\rlap{.}^\circ51$ &
6500-- 7750                                    &
$3\times$16                                    &
1.33                                           &
$1\rlap{.}^{\prime\prime}9$                      \\
24/09/13                                       &
PG0900/(no filter)                             &
$1\rlap{.}^{\prime\prime}00\times8^\prime$        &
1000--1500                                     &
\llap{1}$2\rlap{.}^\circ12$/$24\rlap{.}^\circ25$ &
3800-- 6180                                    &
$3\times$13                                    &
1.32                                           &
$2\rlap{.}^{\prime\prime}0$                      \\
\hline
\end{tabular}
\end{table*}

The observations reported here were obtained with SALT (Buckley, Swart \&
Meiring 2006) under programmes 2012-1-UKSC-002 \& 2013-1-UKSC-007 (PI: van
Loon). With an $11\times10$ m$^2$ segmented spherical primary mirror across
which the pupil is tracked, SALT is the largest single optical telescope in
the world. It is situated at the South African Astronomical Observatory (SAAO)
site in the interior of the Northern Cape province in the Republic of South
Africa, at an altitude of 1800m. We made use of the Robert Stobie Spectrograph
(RSS: Burgh et al.\ 2003; Kobulnicky et al.\ 2003) in its long-slit mode. The
details of the observations are summarised in Table 1. The signal was sampled
by $2\times2$ detector elements upon read-out, to limit electronic noise
whilst adequately sampling the spectral and angular resolution elements. Two
gaps appear in the spectrum because of the physical separation between the
three detectors.

The slit was oriented $32^\circ$ east from north -- roughly aligned with the
{\it Spitzer} InfraRed Spectrograph's short-wavelength slit and perpendicular
to its much wider long-wavelength slit (cf.\ H11) -- so as to overlap with an
anonymous star and a resolved galaxy (narrowly missing its nucleus). The star,
which we name [VS2015]\,J053640.54$-$722615.5 (hereafter ``SAGE0536S''), is
optically brighter than the target AGN and was deemed useful to cancel
telluric absorption in the spectrum of the target. The galaxy,
[GC2009]\,J053642.29$-$722556.6 (hereafter ``SAGE0536G'') was discovered, also
serendipitously, by Gruendl \& Chu (2009) in their search for young stellar
objects in the {\it Spitzer} SAGE-LMC photometric survey (Meixner et al.\
2006). SAGE0536G is bright at UV, mid- and far-IR wavelengths but not
particularly bright at radio frequencies (see Fig.\ 1 in H11, where it is the
brightest object at 8--160 $\mu$m).

All data were processed using the PySALT software suite (Crawford et al.\
2010), as follows: (i) preparation for the subsequent procedures using the
task {\sc saltprepare}; (ii) correction for gain differences between the
amplifiers of the different detectors using the task {\sc saltgain}; (iii)
correction for cross-talk between the amplifiers upon read-out of the
detectors using the task {\sc saltxtalk}; (iv) correction for the electronic
off-set by subtracting the bias level determined from the un-illuminated part
of the detector using the task {\sc saltbias}; (v) cleansing the majority of
cosmic ray imprints using the task {\sc saltcrclean}; (vi) mosaicking into a
single frame using the task {\sc saltmosaic}. We derived a wavelength solution
from the exposure of an argon arc lamp which had been taken immediately after
the science exposure, using the interactive task {\sc specidentify}. This was
then applied to the science frames using the task {\sc specrectify}. In the
case of multiple exposures these were combined using the task {\sc
saltcombine}. The tasks {\sc specsky} and {\sc specextract} were used,
respectively, to subtract background light determined from a patch of sky
adjacent to the object of interest and to extract a weighted average of the
rows on the detector covering the object of interest.

The standard star EG\,21 was observed on the $8^{\rm th}$ of September 2012
with a $4^{\prime\prime}$-wide slit but otherwise identical instrumental set-up.
The data were processed in the same fashion as described previously, before
determining a $3^{\rm rd}$-order polynomial response function using the task
{\sc specsens}. This was then applied to the low-resolution spectra using the
task {\sc speccal}, which also includes a correction for the atmospheric
continuum extinction but not discrete telluric features. The latter were
removed by division by a normalised template incorporating two telluric oxygen
bands as observed in the spectrum of SAGE0536S. Artefacts remaining from
imperfect removal of cosmic rays and telluric emission were removed manually.

%=========================================================================== 3
\section{Scientific results}

%
% FIGURE 1
%
\begin{figure}
\centerline{\psfig{figure=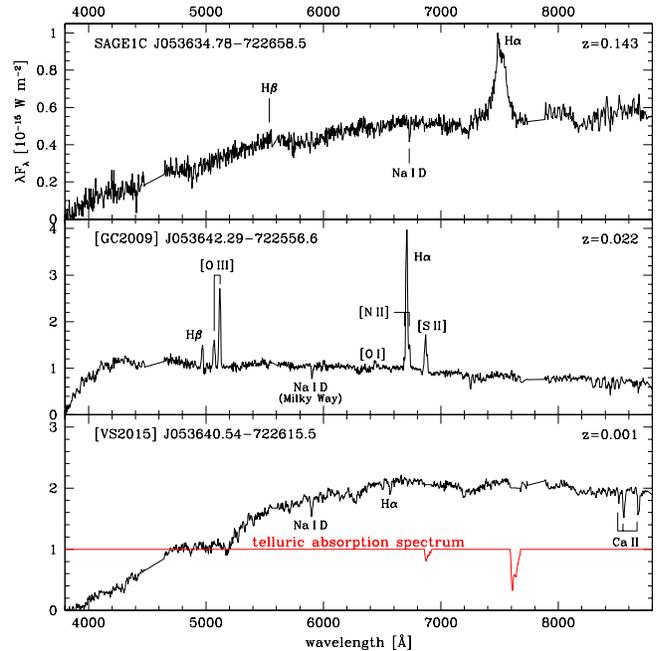,width=85mm}}
\caption[]{SALT-RSS spectra of the ({\it top}) AGN, ({\it middle}) galaxy, and
({\it bottom}) star at low resolution covering most of the optical range.
Conspicuous spectral features are labelled. The spectra have {\it not} been
corrected for redshift.}
\end{figure}

%
% FIGURE 2
%
\begin{figure*}
\centerline{\hbox{
\psfig{figure=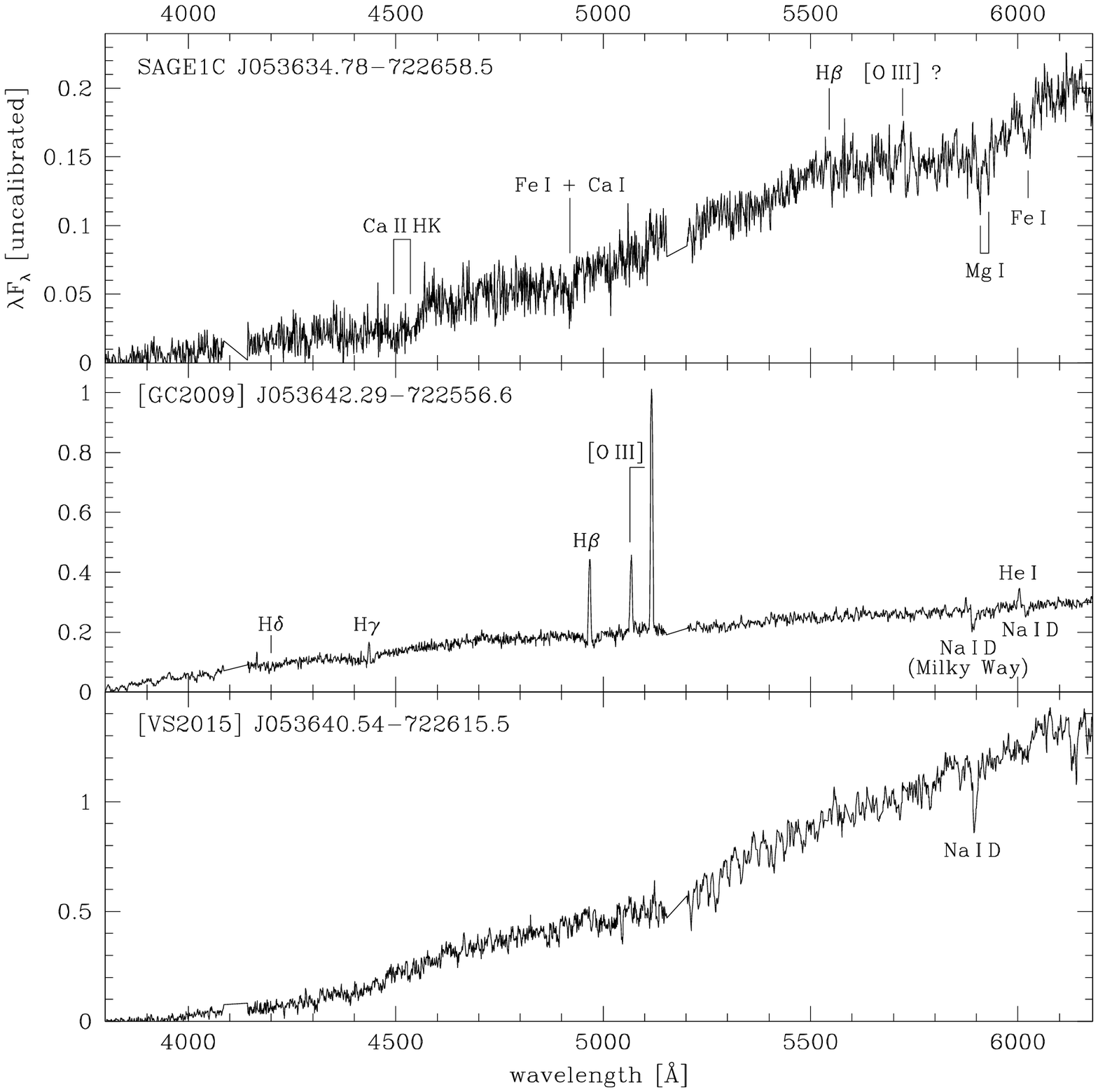,width=88mm}
\psfig{figure=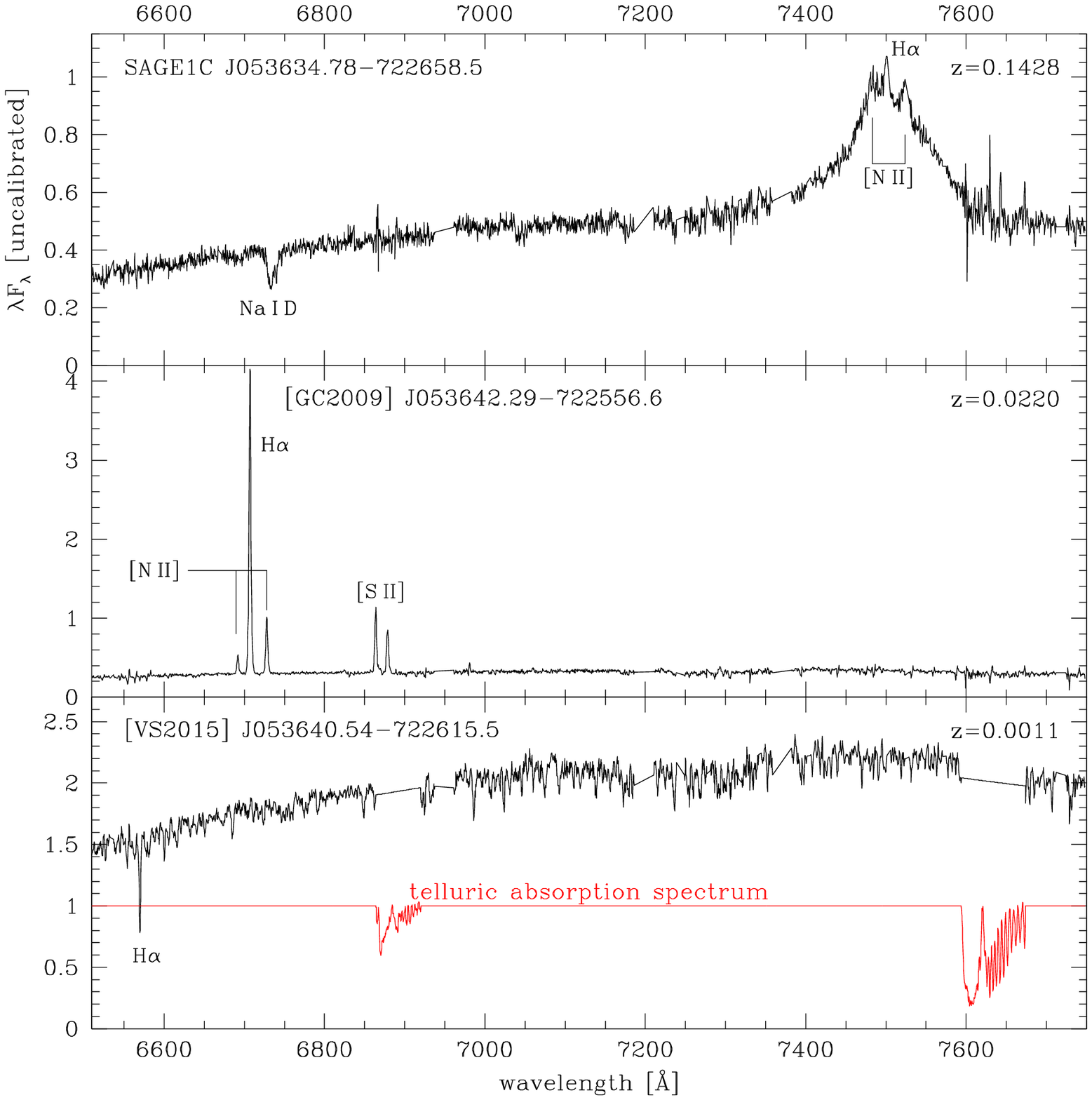,width=88mm}
}}
\caption[]{As Fig.\ 1, but: ({\it left}) at medium resolution covering the
blue--orange and ({\it right}) at high resolution covering the red portion.}
\end{figure*}

%------------------------------------------------------------------------- 3.1
\subsection{Overall spectral appearance: the nature of the AGN host galaxy}

Figures 1 \& 2 show the spectra of SAGE0536AGN, G and S. The first observation
one could make is that SAGE0536AGN exhibits broad H$\alpha$ emission,
SAGE0536G displays narrow Balmer and forbidden line emission, whilst SAGE0536S
only shows absorption lines. Secondly, SAGE0536AGN is redshifted with respect
to SAGE0536G, which is redshifted with respect to SAGE0536S. We thus identify
SAGE0536AGN as an AGN, SAGE0536G as a star-forming galaxy, and SAGE0536S as a
cool star.

The weakness of any narrow emission lines, in particular of [O\,{\sc iii}], in
the spectrum of SAGE0536AGN is of significance as it means there is next to no
current star formation activity. This is consistent with the lack of far-IR
emission (H11), and implies that we are dealing with an early-type galaxy.
Indeed, the lack of strong Balmer absorption lines but presence of metal
absorption lines implies the absence of a sizeable population of stars younger
than a few Gyr. The jump in the spectrum around a restwavelength of 4000 \AA,
$D_n(4000)$ (Balogh et al.\ 1999) and the H$\delta$ line index, $H\delta_A$
(Worthey \& Ottaviani 1997) can be used as age indicators. We estimate
$D_n(4000)\sim1.91$ and $H\delta_A\sim-0.09$ \AA, which (including the
look-back time -- see below) would correspond to a dominant age of $\sim9$
Gyr, or even older for sub-solar metallicity (Kauffmann et al.\ 2003a).

The broad H$\alpha$ emission in SAGE0536AGN is superimposed with weak, narrow
line emission from H$\alpha$ and [N\,{\sc ii}]. The dominance of the BLR over
the NLR suggests that the orientation of SAGE0536AGN is such that we have a
clear view of the central engine, unobstructed by the dusty torus that governs
the IR emission (cf.\ Antonucci \& Miller 1985; Antonucci 1993; Urry \&
Padovani 1995; Beckmann \& Shrader 2012). The narrow H$\alpha$ and [N\,{\sc
ii}] lines are about equally strong; this already suggests an AGN origin
(Baldwin, Phillips \& Terlevich 1981; Kaufmann et al.\ 2003a). While detection
of the [O\,{\sc iii}] line is highly uncertain it may be about twice as bright
as the narrow H$\beta$ line (Fig.\ 2). This would place it among the LINERS
(Kewley et al.\ 2001; Kaufmann et al.\ 2003a).

%------------------------------------------------------------------------- 3.2
\subsection{Kinematics: galaxy redshift, galaxy mass and black hole mass}

The recession velocity of SAGE0536AGN results in a redshift
$z=0.1428\pm0.0001$, which corresponds to a distance of 700 Mpc, k-corrected
distance modulus of 39.08 mag and a look-back time of 1.86 Gyr according to
the standard $\Lambda$-CDM model with Hubble constant $H_0=67.8$ km s$^{-1}$
Mpc$^{-1}$ and matter density $\Omega_0=0.308$ (cf.\ Planck Collaboration
2015). This redshift value is consistent with the estimate from the PAH
features in the mid-IR (H11). This confirms that the optical object is the
same as the IR object. The other galaxy, SAGE0536G has a redshift
$z=0.0220\pm0.0001$ (a distance of 100 Mpc); the line emission shows a
gradient along the slit of $\sim\pm20$ km s$^{-1}$, presumably due to the
(projected) rotation of the disc of this galaxy. The redshift of the star,
SAGE0536S is $z=0.0011\pm0.0001$, which places it within the LMC.

The stellar velocity dispersion in SAGE0536AGN was determined with the {\sc
ppxf} software (Cappellari \& Emsellem 2004) to fit a spectrum of a Lick
standard star (HD\,6203) to the 5695--6070 \AA\ rest wavelength spectral
region. HD\,6203 had been observed using the PG1300 grating but with an
$0\rlap{.}^{\prime\prime}6$ slit and hence a spectral resolution that matches
well that of our PG1800 spectrum. The Na\,{\sc i}\,D is by far the most
conspicuous stellar feature in the spectrum of SAGE0536AGN and drives the
value for the velocity dispersion, $\sigma=123\pm15$ km s$^{-1}$. This makes
it a modest-mass galaxy, and intermediate-mass for an early-type galaxy --
$M_{\rm stars}\sim2.5\times10^{10}$ M$_\odot$ accurate to within about a factor
two based on the uncertainty in the velocity dispersion and spread in the
$M_{\rm stars}$--$\sigma$ relation (Wake, van Dokkum \& Franx 2012; Cappellari
et al.\ 2013; Belli, Newman \& Ellis 2014). The visual brightness is $V\sim18$
mag (estimated from the B- and i-band photometry in H11), hence
$L(V)\sim2\times10^{10}$ L$_\odot$.

As motions within the BLR are dominated by the gravitational well of the
central black hole, the H$\beta$ line profile yields an estimate of the black
hole mass (Kaspi et al.\ 2000; Vestergaard \& Peterson 2006). Because of the
faintness of the H$\beta$ emission we use instead the H$\alpha$ line, which
has a Full Width at Half Maximum (FWHM) $\sim4000\pm400$ km s$^{-1}$
(corresponding to a Gaussian $\sigma\sim1700$ km s$^{-1}$). Following Woo et
al.\ (2014), who present a relation for the black hole mass derived solely
from the H$\alpha$ line profile for type I AGNs:
\begin{equation}
\frac{M_{\rm BH}}{{\rm M}_\odot}\ =\ 2.2\times10^7\times
\left[\frac{\sigma({\rm H}\alpha)}{10^3\,{\rm km\,s}^{-1}}\right]^{2.06}
\left[\frac{L({\rm H}\alpha)}{10^{42}\,{\rm erg\,s}^{-1}}\right]^{0.46}
\end{equation}
The H$\alpha$ luminosity is estimated from the integrated line profile and the
i-band magnitude (H11) -- $L({\rm H}\alpha)\sim4\pm0.4\times10^{43}$ erg
s$^{-1}$. Hence, $M_{\rm BH}\sim3.5\pm0.8\times10^8$ M$_\odot$.

%------------------------------------------------------------------------- 3.3
\subsection{Foreground interstellar absorption}

The spectrum of SAGE0536G shows absorption in the Na\,{\sc i}\,D doublet at
5890+5896 \AA, arising within the interstellar medium of the Milky Way and/or
LMC (van Loon et al.\ 2013); foreground absorption in the diffuse interstellar
band at 4428 \AA\ is uncertain as it would coincide with redshifted Br$\gamma$
arising in SAGE0536G. Likewise, any foreground Na\,{\sc i}\,D absorption in
the spectrum of SAGE0536AGN would coincide with redshifted Mg\,{\sc i} whilst
in the spectrum of SAGE0536S it would coincide with the stellar photospheric
Na\,{\sc i}\,D absorption.

%=========================================================================== 4
\section{Discussion on the nature of SAGE0536AGN}

%------------------------------------------------------------------------- 4.1
\subsection{The host galaxy}

The brightness of SAGE0536AGN at radio, compared to that at optical
frequencies, $F_\nu(4.75\,{\rm GHz})/F_\nu({\rm B})=40\,{\rm mJy}/0.162\,{\rm
mJy}\gg 10$ makes it a radio-loud AGN (Kellermann et al.\ 1989). On the other
hand, $L_\nu(178\,{\rm MHz})\sim4\times10^{21}\ll 2\times10^{25}$ W Hz$^{-1}$
sr$^{-1}$ places it in the Fanaroff--Riley class I (Fanaroff \& Riley 1974);
SAGE0536AGN may well have the bright radio lobes that are characteristic of
radio-loud galaxies, but if viewed close to pole-on it may resemble a compact
FR-I type radio source.

The faintness at optical wavelengths ($M_{\rm B}=-20.6$ mag) and brightness at
X-ray frequencies ($L_{\rm X}(0.1$--$2.4\,{\rm keV})=1.5\times10^{43}$ erg
s$^{-1}$ -- see H11), and relative strength of broad compared to narrow Balmer
lines, places SAGE0536AGN in the category of Seyfert 1 (Schmidt \& Green 1983;
Osterbrock 1977, 1981). While elliptical galaxies are the typical host of a
broad-line radio AGN, most Seyferts are radio-quiet AGN residing in spiral
galaxies. There thus seems to be some ambiguity as to the expected host galaxy
type of SAGE0536AGN, which we have shown here is of early-type and definitely
not a spiral galaxy. Indeed, in the diagnostic diagram of H$\alpha$ equivalent
width versus [N\,{\sc ii}]/H$\alpha$ ratio -- which Cid Fernandes et al.\
(2010) introduced to circumvent problems with weak or absent H$\beta$ and/or
[O\,{\sc iii}] lines which traditionally form part of line diagnostic diagrams
-- SAGE0536AGN has much weaker [N\,{\sc ii}] than is usual for Seyferts.
However, this is probably because we are comparing the [N\,{\sc ii}] from the
NLR with the H$\alpha$ from the BLR.

The UV-to-IR ratio is different for early- and late-type galaxies. SAGE0536AGN
was not detected at near-UV (NUV) wavelengths with GALEX, however, to a limit
of 3 $\mu$Jy (H11). Following Fukugita et al.\ (1996) this corresponds to an
AB magnitude of 7.07, and hence the NUV--[3.6] colour is $>5$ mag. According
to Bouquin et al.\ (2015) this makes SAGE0536AGN most likely an elliptical or
S0 galaxy. So this is consistent with our optical spectroscopic determination
of the galaxy type.

%
% TABLE 2
%
\begin{table}
\caption{WISE photometry for SAGE0536AGN (from Cutri et al.\ 2014); magnitudes
are on the Vega system and flux densities are isophotal assuming
$F_\nu\propto\nu^{-2}$ and have not been colour-corrected.}
\begin{tabular}{llll}
\hline\hline
band & $\lambda$ ($\mu$m) & magnitude               & $F_\nu$ (mJy) \\
\hline
W1   &         3.4        & \llap{1}2.404$\pm$0.024 &         3.4  \\
W2   &         4.6        & \llap{1}1.386$\pm$0.021 &         4.8  \\
W3   & \llap{1}2          &         8.610$\pm$0.020 & \llap{1}1.4  \\
W4   & \llap{2}2          &         6.607$\pm$0.043 & \llap{1}9.0  \\
\hline
\end{tabular}
\end{table}

Near-IR colours are not very discriminative, with the $J-K_{\rm s}\sim2$ mag of
SAGE0536AGN placing it among moderately obscured AGN that include a wide range
of host galaxy and AGN types (Rose et al.\ 2013). Mid-IR colours, however, can
distinguish between different kinds of objects, as LaMassa et al.\ (2013b)
have demonstrated using WISE data. SAGE0536AGN was in fact detected in all
four WISE bands (Table 2; Cutri et al.\ 2014). According to Fig.\ 11 in
LaMassa et al.\ (2013b) and Fig.\ 5 in Cluver et al.\ (2014) this firmly
places SAGE0536AGN among the QSOs, i.e.\ a galaxy dominated by its AGN.
Indeed, the mid-IR flux densities roughly follow a powerlaw
$F_\nu\propto\nu^{-0.92}$ rather than a black-body (cf.\ Yang, Chen \& Huang
2015). The dominance of the AGN over emission from the host galaxy may be a
reason why the usual diagnostics based on optical, X-ray and radio data yield
ambiguous expectations for the galaxy type of SAGE0536AGN. Even mid-IR
diagrams are not without ambiguity; in the WISE [4.6]--[12] versus galaxy mass
diagram of Alatalo et al.\ (2014) the sequences of early-type (blue
[4.6]--[12] colours) and late-type (red [4.6]--[12] colours) galaxies separate
very well yet SAGE0536AGN would clearly -- and mistakenly -- end up among the
late-type galaxies (and optical `green valley' galaxies) because its AGN
dominates over the host galaxy.

The lack of far-IR emission places stringent constraints, not only on the star
formation rate but also on the mass of interstellar gas. Groves et al.\ (2015)
found relations between the total interstellar hydrogen mass and the
luminosity in {\it Herschel} Space Observatory bands, for a sample of mostly
irregular and spiral galaxies. The strongest constraint on the gass mass in
SAGE0536AGN is set by the 160-$\mu$m 3-$\sigma$ upper limit of 4 mJy (H11),
yielding $M_{\rm gas}<10^9$ M$_\odot$ (according to Groves et al.\ 2015). With
$M_{\rm gas}/M_{\rm stars}<0.1$ this implies it is gas-poor, as expected for an
early-type galaxy. The corresponding limit on the star formation rate is
(Willott, Bergeron \& Omont 2015)
$<1.5\times10^{-10}L_{\rm FIR}({\rm L}_\odot)\sim0.26$ M$_\odot$ yr$^{-1}$ or a
specific star formation rate $<1.3\times10^{-11}$ yr$^{-1}$ -- clearly below
that on the main sequence of star-forming galaxies (Wuyts et al.\ 2012;
Vulcani et al.\ 2015; Renzini \& Peng 2015).

The narrow H$\alpha$ emission component has an equivalent width $\sim1.7$ \AA\
(accurate to $\sim10$ per cent). Following Hopkins et al.\ (2003), for an
R-band magnitude $\approx17.5$ (estimated from the B-band and i-band
photometry in H11) and the measured Balmer decrement (see below) and redshift,
this would yield $L_{{\rm H}\alpha}\sim3.4\times10^{35}$ W (accurate to within
about a factor two). If this were due to star formation activity then,
following Wijesinghe et al.\ (2011), the star formation rate would be $\sim10$
M$_\odot$ yr$^{-1}$. The fact that this is forty times above the upper limit
derived from the far-IR luminosity means that the narrow H$\alpha$ emission
component must be associated with the NLR of the AGN instead.

%------------------------------------------------------------------------- 4.2
\subsection{The central engine}

%
% FIGURE 3
%
\begin{figure}
\centerline{\psfig{figure=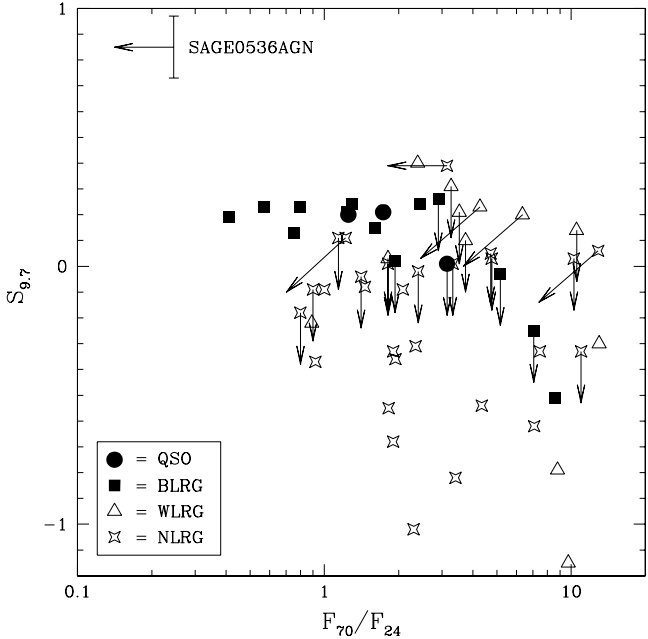,width=85mm}}
\caption[]{Silicate strength $S_{9.7}$ versus $F_{70}/F_{24}$ flux ratio for
narrow-line, weak-line and broad-line radio galaxies and QSOs (Dicken et al.\
2008, 2010, 2014) and SAGE0536AGN.}
\end{figure}

SAGE0536AGN stands out by its extremely strong silicate emission at 9.7
$\mu$m. Strong silicate emission is typical of broad-line radio galaxies
(Dicken et al.\ 2014), so our BLR-dominated optical spectrum is consistent
with its known radio loudness (H11). However, while its $F_{20}/F_7$ flux
ratio (estimated from the {\it Spitzer} spectrum at 7 and 20 $\mu$m) is not
unusual for this kind of galaxy/AGN, it takes an extreme position in a diagram
of silicate strength ($S_{9.7}=\ln(F_{9.7}/F_{\rm continuum})$ -- Spoon et al.\
2007) versus $F_{70}/F_{24}$ flux ratio (Fig.\ 3) compared to a sample of radio
galaxies with {\it Spitzer} spectroscopy (Dicken et al.\ 2014). SAGE0536AGN
shares the characteristics of strong silicate emission and low $F_{70}/F_{24}$
flux ratio with BLR galaxies, which is consistent with its optical spectrum
being dominated by BLR emission. however, it is more extreme than the BLR
galaxies featuring in this plot both in its silicate emission {\it and} in its
$F_{70}/F_{24}$ flux ratio. It is possible that this is related to its modest
host galaxy mass, compared to that of the comparison sample of BLR galaxies.

Gandhi et al.\ (2009) have shown that an excellent relation exists between the
spatially resolved mid-IR luminosity of a Seyfert-type AGN and its X-ray
luminosity. We could consider whether the same relationship might also apply
to SAGE0536AGN. With an estimated foreground (Galactic + LMC) hydrogen column
density of $N({\rm H})\approx 1\times10^{21}$ cm$^{-2}$ (Staveley-Smith et al.\
2003), and assuming a photon index of 2 (P.Gandhi, private communication) we
use the ROSAT measurement (see H11) and the PIMMS
tool\footnote{https://heasarc.gsfc.nasa.gov/cgi-bin/Tools/w3pimms/w3pimms.pl}
to estimate that $L_{\rm X}(2$--$10\,{\rm keV})\sim1.8\times10^{43}$ erg
s$^{-1}$. According to the relation in Gandhi et al.\ (2009)
\begin{equation}
\log\left[\frac{L_{\rm IR}(12.3\,\mu{\rm m})}{10^{43}\,{\rm erg\,s}^{-1}}\right]
=0.19+1.11\log\left[
\frac{L_{\rm X}(2{\rm -}10\,{\rm keV})}{10^{43}\,{\rm erg\,s}^{-1}}\right]
\end{equation}
this would predict $L_{\rm IR}(12.3\,\mu{\rm m})\sim3\times10^{43}$ erg
s$^{-1}$. The observed mid-IR luminosity of SAGE0536AGN is
$L_{\rm IR}(12.3\,\mu{\rm m})\sim1.3\times10^{44}$ erg s$^{-1}$, i.e.\ a factor
$\sim4$ brighter. We could reconcile the two if we added internal absorption
equivalent to $N({\rm H})\sim2\times10^{22}$ cm$^{-2}$. This neutral gas, along
with the dust may be a wind seen in front of the nuclear engine (cf.\ H\"onig
et al.\ 2013).

Indeed, the H$\alpha$/H$\beta$ peak ratio in our spectra is $\sim20$ and hence
$\gg 2.8$ (as expected for Case B recombination; see Osterbrock 1974) and thus
suggestive of differential extinction (reddening) by dust. A factor 7
differential extinction at these wavelengths corresponds to $E(B-V)\sim1.6$
mag (cf.\ Lyu, Hao \& Li 2014). With $N({\rm H})\approx 2.2\times10^{21}A_V$
cm$^{-2}$ (G\"uver \& \"Ozel 2009) and a canonical $A_V\approx3.1E(B-V)$ this
would suggest an associated hydrogen column density
$N({\rm H})\sim 1\times10^{22}$ cm$^{-2}$. The factor two difference with the
above estimate based on the X-ray attenuation can easily be accounted for by
lowering the dust:gas ratio by a factor two compared to what is typical in the
neutral ISM of the Milky Way, for instance if the gas is metal-poor (Tatton et
al.\ 2013) or if it sublimates in the vicinity of the AGN. Thus, while each of
the above determinations carries uncertainties in measurement and assumption,
the fact that these independent measurements (X-ray, IR, optical lines) can be
understood within a single, consistent picture lends credibility to the
presence of a dusty X-ray absorber near to the central engine in SAGE0536AGN.

%
% FIGURE 4
%
\begin{figure}
\centerline{\psfig{figure=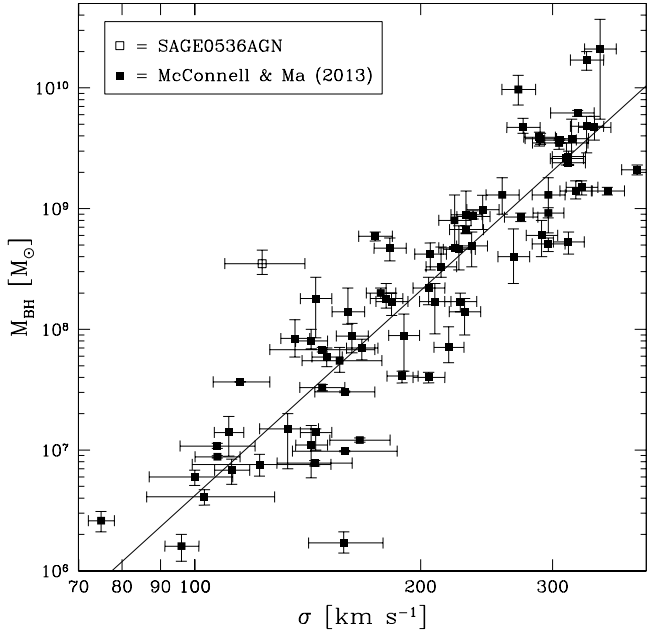,width=85mm}}
\caption[]{Black hole mass versus velocity dispersion. The McConnell \& Ma
(2013) sample are shown by filled symbols and their best-fit relation; their
sample is a mixture of early- and late-type galaxies. SAGE0536AGN, indicated
with the open symbol, is a clear outlier towards high black hole mass.}
\end{figure}

With a host galaxy stellar velocity dispersion of $\sigma=123$ km s$^{-1}$ we
would have expected $M_{\rm BH}\sim1.8\times10^7$ M$_\odot$ (e.g., Ferrarese \&
Merritt 2000; Gebhardt et al.\ 2000; Tremaine et al.\ 2002; G\"ultekin et al.\
2009; McConnell \& Ma 2013). On the basis of the estimated host galaxy mass we
would have expected $M_{\rm BH}\sim5\times10^7$ M$_\odot$ (Sanghvi et al.\
2014). These are an order of magnitude below the
$M_{\rm BH}=3.5\pm0.8\times10^8$ M$_\odot$ we determined for SAGE0536AGN; in
figure 4 we plot the location of SAGE0536AGN in a $M_{\rm BH}$--$\sigma$
diagram along with the data and best fit line from McConnell \& Ma (2014). The
latter consists of a mix of early- and late-type galaxies, but they do not
differ much in their occupation of the diagram. Admittedly, the scatter around
these relations is significant (Salviander \& Shields 2013). The fact that
SAGE0536AGN lies within a factor two of the $M_{\rm BH}$--$M_{\rm spheroid}$
(Graham \& Scott 2015) and $M_{\rm BH}$--$L_{\rm spheroid}(V)$ (McConnell \& Ma
2013; Park et al.\ 2015) relations may be regarded as confirmation that the
host galaxy of SAGE0536AGN is indeed dominated by a spheroidal component.
Over-massive black holes have been suggested in other galaxies, but these are
based on luminosity determinations of the galaxy mass which depend on galaxy
morphology; in the $M_{\rm BH}$--$\sigma$ diagram they occupy unremarkable
positions (Kormendy \& Ho 2013).

McQuillin \& McLaughlin (2013) explain the $M_{\rm BH}$--$\sigma$ relation as
resulting from self-regulated feedback, in which the black hole outflow speed,
$v_{\rm outflow}$ features: $M_{\rm BH}v_{\rm outflow}\propto\sigma^5$ and thus any
black hole outflow in SAGE0536AGN is expected to be weak,
$v_{\rm outflow}\sim0.002$ c compared to a median of $v_{\rm outflow}\sim0.035$ c
for local AGN (McQuillin \& McLaughlin 2013). A slow outflow would also be
consistent with the relatively high (for an AGN seen pole-on) hydrogen and
dust column density as estimated above. Stronger winds in starbursts sometimes
give rise to Na\,{\sc i}\,D absorption (Rupke et al.\ 2005) but SAGE0536AGN is
not a starburst galaxy and its Na\,{\sc i}\,D profile shows no sign of
significant broadening or displacement as one may expect from a strong wind.

The luminosity of the AGN can be determined from the H$\alpha$ luminosity (Woo
et al.\ 2014):
\begin{equation}
\frac{L_{\rm bol}({\rm AGN})}{10^{44}\,{\rm erg\,s}^{-1}}=2.21\times
\left[\frac{L({\rm H}\alpha)}{10^{42}\,{\rm erg\,s}^{-1}}\right]^{0.86}
\end{equation}
Hence we obtain $L_{\rm bol}({\rm AGN})=5.3\pm0.4\times10^{45}$ erg s$^{-1}$.
With an Eddington luminosity of $L_{\rm edd}=1.25\times10^{38}M({\rm BH})$ erg
s$^{-1}$ (Wyithe \& Loeb 2002) we obtain
$L_{\rm bol}({\rm AGN})/L_{\rm edd}=0.12$. Hence the AGN in SAGE0536AGN is as
active as one may expect for a black hole of this size -- at high $M_{\rm BH}$
and $L_{\rm bol}({\rm AGN})$ where almost all AGN are of optical type I (Oh et
al.\ 2015).

Interestingly, the extreme silicate strength, moderately strong X-ray absorber
and moderately high accretion rate are in fair agreement with models for an
AGN comprising of a geometrically thin optically thick disc, geometrically
thicker optically thinner medium (e.g., a disc wind) and an outflow cone for
an inclination angle close to zero degrees (Stalevski et al.\ 2012; Schartmann
et al.\ 2014).

%------------------------------------------------------------------------- 4.3
\subsection{An evolutionary picture}

How can we explain the extreme properties of SAGE0536AGN? The
modest star formation rate is typical of a radio-loud galaxy (G\"urkan et al.\
2015). Given the
predominance of the BLR we could be looking into the heart of the AGN. The host is a rather modest-mass galaxy -- not a giant cD elliptical
galaxy such as M\,87 of which the AGN is fed by cooling flows from the cluster
inter-galactic medium. But SAGE0536AGN hosts a rather massive and actively
accreting nuclear black hole. While the evidence we present here is in favour
of a predominantly pressure-supported mature stellar system, we cannot fully
rule out a lenticular galaxy with some rotational support, seen face-on --
this could lower the observed velocity dispersion and bring it closer to the
canonical $M_{\rm BH}$--$\sigma$ relation.

The black hole in SAGE0536AGN could have been fed recently by a gas-rich minor
merger or cooling flows from its circum-galactic environment. This would
explain its current level of AGN activity for such modest galaxy host. Traces
of intermediate-age populations in early-type galaxies of QSOs at redshifts
$z\sim0.2$ have recently been suggested to represent the aftermath of mergers
(Canalizo \& Stockton 2013), but no such intermediate-age populations are
detected in SAGE0536AGN. Minor mergers have also been invoked to explain dust
lanes in early-type galaxies, in which the star formation efficiency is very
low (Davis et al.\ 2015). A merger scenario does not offer an explanation for
the exuberant mass of the black hole, but black holes are in fact seen to grow
at a faster rate than the galaxies they inhabit (Diamond-Stanic \& Rieke 2012;
LaMassa et al.\ 2013a), and galaxy growth may even be prematurely curtailed
(Ferr\'e-Mateu et al.\ 2015).

Nature appears to fine-tune the competition between circum-galactic cooling
flows feeding a nuclear black hole or feeding star formation within the host
galaxy, and the feedback from the black hole (and possibly star formation)
heating the accreted gas thus preventing stars from forming. This balance
relies critically on the timescales and development of thermal instabilities
within the gas (Voit et al.\ 2015), and it may have shifted in the case of
SAGE0536AGN towards a relatively warm accretion flow and/or rapid feedback
thus leading to black hole growth without the associated stellar mass
build-up.

The AGN activity may now have largely subsided due to a lack of fresh fuel,
having quenched all star formation in its host (cf.\ Stasi\'nska et al.\ 2008;
Cid Fernandes et al.\ 2010; Cimatti et al.\ 2013; Tasca et al.\ 2014; Wong et
al.\ 2015). The black hole mass is now as massive as it gets, having devoured
much of its host's gas. It may resemble a lower-mass counterpart to the
redshift 1.6 radio galaxy 5C\,7.245 (Humphrey et al.\ 2015) or redshift 3.3
galaxy CID-947 (Trakhtenbrot et al.\ 2015); the lower redshift at which we see
something similar happen to SAGE0536AGN is in agreement with the `downsizing'
scenario in which lower-mass galaxies evolve more slowly. Galaxies with
$M_{\rm stars}\sim2.5\times10^{10}$ M$_\odot$ are indeed seen to be nearing the
end of their assembly history around redshift $z\sim0.1$--0.2 (McDermid et
al.\ 2015). However, the old age of SAGE0536AGN ($>9$ Gyr, i.e.\ $z>2.25$)
seems incompatible with this evolutionary scenario. Interestingly, some nearby
star-forming dwarf galaxies with broad-line AGN have been found with
$M_{\rm BH}$ about seven times that expected from the
$M_{\rm BH}$--$M_{\rm stars}$ relation (viz.\ galaxies B and D in Reines, Greene
\& Geha 2013). The latter could represent even more delayed transformation of
even lower-mass AGN-host galaxies, again consistent with the mass-dependent
galaxy assembly history paradigm (McDermid et al.\ 2015). On the other hand,
this would fail to explain the over-massive black hole in the nearby {\it
massive} lenticular galaxy NGC\,1277 (Scharw\"achter et al.\ 2015).

The effective radius $R_{\rm e}$ of the host galaxy may be the parameter which
improves upon the $M_{\rm BH}$--$\sigma$ relation to create a fundamental plane
of $M_{\rm BH}$--$\sigma$--$R_{\rm e}$ (Marconi \& Hunt 2003; de Francesco,
Capetti \& Marconi 2006). If SAGE0536AGN is in transformation from a late-type
to an early-type galaxy, its effective radius may now be relatively large.
This would be expected for newly quenched early-type galaxies (Daddi et al.\
2005; Trujillo et al.\ 2007; Carollo et al.\ 2013; Cassata et al.\ 2013;
Poggianti et al.\ 2013), but SAGE0536AGN appears to have stopped forming stars
in abundance already some 9 Gyr ago.

Much to the contrary, Katkov, Kniazev \& Sil'chenko (2015) suggest lenticular
galaxies are only now building up discs that may see them eventually transform
into spiral galaxies. The mechanism for this transformation is accretion of
cold gas from their surroundings, which could indeed fuel also a nuclear black
hole. The large spread in ages for lenticular (S0) galaxies (Poggianti et al.\
2001, 2009) -- which have $\sigma$ broadly similar to that of SAGE0536AGN --
indicates that this process may still be on-going at $z<1$. While it does not
explain the large black hole mass in SAGE0536AGN, aside from its low $\sigma$
possibly being due to projection effects it could also mean that SAGE0536AGN
may evolve closer to the canonical $M_{\rm BH}$--$\sigma$ relation as the
galaxy grows in mass (cf.\ DeGraf et al.\ 2015). On the other hand, there is
no evidence of stars younger than $\sim9$ Gyr in the optical spectrum of
SAGE0536AGN, which seems to be in tension with it building up a disc now.

An alternative explanation for super-massive black holes was offered by
Volonteri \& Ciotti (2013) for central cluster galaxies, which they suggest
are predominantly the result from dry mergers that do little to increase the
velocity dispersion. Savorgnan \& Graham (2015) found a lack of evidence for
this scenario among massive galaxies. That does not mean that it may not be a
viable option for less massive galaxies such as SAGE0536AGN. It could explain
why it lies closer to the $M_{\rm BH}$--$M_{\rm spheroid}$ relation than to the
$M_{\rm BH}$--$\sigma$ relation.

There must be other objects like SAGE0536AGN, and these could well form a new
sub-class of weak-line radio galaxies (WLRG) with warm dust. WLRG galaxies are
defined by their optical [O\,{\sc iii}] equivalent width $<10$ \AA\ (Tadhunter
et al.\ 1998); they are not normally associated with dust and it has therefore
been suggested that they may be accreting directly from the hot inter-galactic
medium (Hardcastle, Evans \& Croston 2007). SAGE0536AGN would classify as a
WLRG, but it does contain dust. Another WLRG, PKS\,0043$-$42 at $z=0.116$
(Tadhunter et al.\ 1993) also exhibits warm dust, though with
$F_{70}/F_{24}=0.9$ it is less extreme than SAGE0536AGN. While in
PKS\,0043$-$42 the silicate is seen in absorption (Ramos Almeida et al.\ 2011)
and it is a FR-II object, these differences with SAGE0536AGN could be largely
explained by orientation effects with PKS\,0043$-$42 being seen edge on.

%=========================================================================== 5
\section{Conclusions}

Optical spectra obtained with the SALT have allowed us to better characterise
the peculiar AGN-dominated galaxy SAGE1C\,J053634.78$-$722658.5 (SAGE0536AGN
for short):
\begin{itemize}
\item[$\bullet$]{We determine an accurate redshift of $z=0.1428\pm0.0001$ from
the optical absorption and emission lines.}
\item[$\bullet$]{We confirm the strong presence of an AGN, of type I based on
broad H$\alpha$ emission.}
\item[$\bullet$]{We determine the black hole mass
$M_{\rm BH}\sim3.5\pm0.8\times10^8$ M$_\odot$, on the basis of the H$\alpha$
width and luminosity, accreting at $\sim12$ per cent of the Eddington rate.}
\item[$\bullet$]{Narrower, weaker H$\alpha$ and [N\,{\sc ii}] emission is
attributed to the NLR around the AGN; limits on the far-IR emission rule out a
star formation origin of the line emission.}
\item[$\bullet$]{The overall appearance of the optical spectrum is that of an
early-type galaxy at least $\sim9$ Gyr old, consistent with the absence of
star formation indicators.}
\item[$\bullet$]{We measure a stellar velocity dispersion $\sigma=123\pm15$ km
s$^{-1}$ and hence derive a host galaxy mass
$M_{\rm stars}\sim2.5^{2.5}_{1.2}\times10^{10}$ M$_\odot$.}
\item[$\bullet$]{The mid-IR properties, in particular the intense silicate
emission, point towards an AGN torus seen close to pole-on but with moderate
X-ray absorption possibly associated with a dusty wind.}
\item[$\bullet$]{SAGE0536AGN lies (well) above the canonical
$M_{\rm BH}$--$\sigma$ and $M_{\rm BH}$--$M_{\rm stars}$ relations, explaining
its original discovery by Hony et al.\ (2011) as an AGN-dominated galaxy.}
\end{itemize}
Outliers like SAGE0536AGN may provide useful clues as to how the observed
scaling relations arise and evolve. SAGE0536AGN will need to accumulate
stellar mass in future, if it is going to move closer to the main scaling
relations between $M_{\rm BH}$ and host galaxy properties. But it does not
appear to have an internal reservoir of gas to do so.

Deep, high-resolution optical images of the host galaxy are needed to confirm
its morphological type and to determine the effective radius -- to improve the
determination of the mass of the spheroidal component -- thus elucidating the
nature and evolutionary status of this remarkable galaxy.

%=============================================================================
\section*{Acknowledgments}

We would like to thank Poshak Gandhi for discussions regarding the X-ray data,
and the referee, Daniel Dicken, for a positive report and queries which have
helped us to clarify some of our reasoning. All of the observations reported
in this paper were obtained with the Southern African Large Telescope (SALT).
Keele University and the University of Central Lancashire form part of the
United Kingdom SALT Consortium, a partner in SALT. This publication makes use
of data products from the Wide-field Infrared Survey Explorer, which is a
joint project of the University of California, Los Angeles, and the Jet
Propulsion Laboratory/California Institute of Technology, funded by the
National Aeronautics and Space Administration. This research has made use of
the SIMBAD database, operated at CDS, Strasbourg, France.

%=============================================================================

\label{lastpage}


\begin{thebibliography}{}
\bibitem[Alatalo et al.(2014)]{Alatalo14}
Alatalo K., Cales S.\ L., Appleton P.\ N., Kewley L.\ J., Lacy M., Lisenfeld
U., Nyland K., Rich J.\ A., 2014, ApJ, 794, L13
\bibitem[Antonucci (1993)]{Antonucci93}
Antonucci R., 1993, ARA\&A, 31, 473
\bibitem[Antonucci \& Miller (1985)]{Antonucci85}
Antonucci R.\ R.\ J., Miller J.\ S., 1985, ApJ, 297, 621
\bibitem[Baldwin, Phillips \& Terlevich (1981)]{Baldwin81}
Baldwin J.\ A., Phillips M.\ M., Terlevich R., 1981, PASP, 93, 5
\bibitem[Balogh et al.(1999)]{Balogh99}
Balogh M.\ L., Morris S.\ L., Lee H.\ K.\ C., Carlberg R.\ G., Ellingson E.,
1999, ApJ, 527, 54
\bibitem[Beckmann \& Shrader (2012)]{Beckmann12}
Beckmann V., Shrader C.\ R., 2012, Active Galactic Nuclei, Wiley--VCH Verlag
GmbH
\bibitem[Belli, Newman \& Ellis (2014)]{Belli14}
Belli S., Newman A.\ B., Ellis R.\ S., 2014, ApJ, 783, 117
\bibitem[Bouquin et al.(2015)]{Bouquin15}
Bouquin A.\ Y.\ K., 2015, ApJ, 800, L19
\bibitem[Buckley, Swart \& Meiring (2006)]{Buckley06}
Buckley D.\ A.\ H., Swart G.\ P., Meiring J.\ G., 2006, SPIE, 6267, 32
\bibitem[Burgh et al.(2003)]{Burgh03}
Burgh E.\ B., Nordsieck K.\ H., Kobulnicky H.\ A., Williams T.\ B., O'Donoghue
D., Smith M.\ P., Percival J.\ W., 2003, SPIE, 4841, 1463
\bibitem[Canalizo \& Stockton (2013)]{Canalizo13}
Canalizo G., Stockton A., 2013, ApJ, 772, 132
\bibitem[Cappellari \& Emsellem (2004)]{Cappellari04}
Cappellari M., Emsellem E., 2004, PASP, 116, 138
\bibitem[Cappellari et al.(2013)]{Cappellari13}
Cappellari M., et al., 2013, MNRAS, 432, 1862
\bibitem[Carollo et al.(2013)]{Carollo13}
Carollo C.\ M., et al., 2013, ApJ, 773, 112
\bibitem[Cassata et al.(2013)]{Cassata13}
Cassata P., et al., 2013, ApJ, 775, 106
\bibitem[Cid Fernandes et al.(2010)]{Cidfernandes10}
Cid Fernandes R., Stasi\'nska G., Schlickmann M.\ S., Mateus A., Vale Asari
N., Schoenell W., Sodr\'e L., 2010, MNRAS, 403, 1036
\bibitem[Cimatti et al.(2013)]{Cimatti13}
Cimatti A., et al., 2013, ApJ, 779, L13
\bibitem[Cluver et al.(2014)]{Cluver14}
Cluver M.\ E., et al., 2014, ApJ, 782, 90
\bibitem[Crawford et al.(2010)]{Crawford10}
Crawford S.\ M., et al., 2010, SPIE, 7737, 82
\bibitem[Cutri et al.(2014)]{Cutri14}
Cutri, R.\ M., et al., 2014, VizieR On-line Data Catalog: II/328
\bibitem[Daddi et al.(2005)]{Daddi05}
Daddi E., et al., 2005, ApJ, 626, 680
\bibitem[Davis et al.(2015)]{Davis15}
Davis T.\ A., et al., 2015, MNRAS, 449, 3503
\bibitem[de Francesco, Capetti \& Marconi (2006)]{Defrancesco06}
de Francesco G., Capetti A., Marconi A., 2006, A\&A, 460, 439
\bibitem[DeGraf et al.(2015)]{Degraf15}
DeGraf C., Di Matteo T., Treu T., Feng Y., Woo J.-H., Park D., arXiv:1412.4133
\bibitem[Diamond-Stanic \& Rieke (2012)]{Diamondstanic12}
Diamond-Stanic A.\ M., Rieke G.\ H., 2012, ApJ, 746, 168
\bibitem[Dicken et al.(2008)]{Dicken08}
Dicken D., Tadhunter C., Morganti R., Buchanan C., Oosterloo T., Axon D.,
2008, ApJ, 678, 712
\bibitem[Dicken et al.(2010)]{Dicken10}
Dicken D., Tadhunter C., Axon D., Robinson A., Morganti R., Kharb P., 2010,
ApJ, 722, 1333
\bibitem[Dicken et al.(2014)]{Dicken14}
Dicken D., et al., 2014, ApJ, 788, 98
\bibitem[Fanaroff \& Riley (1974)]{Fanaroff74}
Fanaroff B.\ L., Riley J.\ M., 1974, MNRAS, 167, P31
\bibitem[Ferrarese \& Merritt (2000)]{Ferrarese00}
Ferrarese F., Merritt D., 2000, ApJ 539, L9
\bibitem[Ferre\'e-Mateu et al.(2015)]{Ferremateu15}
Ferr\'e-Mateu A., Mezcua M., Trujillo I., Balcells M., van den Bosch R.\ C.\
E., arXiv:1506.02663
\bibitem[Fukugita et al.(1996)]{Fukugita96}
Fukugita M., Ichikawa T., Gunn J.\ E., Doi M., Shimasaku K., Schneider D.\ P.,
1996, AJ, 111, 1748
\bibitem[Gandhi et al.(2009)]{Gandhi09}
Gandhi P., Horst H., Smette A., H\"onig S., Comastri A., Gilli R., Vignali C.,
Duschl W., 2009, A\&A, 502, 457
\bibitem[Gebhardt et al.(2000)]{Gebhardt00}
Gebhardt K., et al., 2000, ApJ, 539, L13
\bibitem[Graham (2015)]{Graham15}
Graham A.\ W., 2015, in: ``Galactic Bulges'', eds.\ E.\ Laurikainen, R.\ F.\
Peletier \& D.\ A.\ Gadotti (Springer), arXiv:1501.02937
\bibitem[Graham \& Scott (2015)]{Graham15}
Graham A.\ W., Scott N., 2015, ApJ, 798, 54
\bibitem[Groves et al.(2015)]{Groves15}
Groves B.\ A., et al., 2015, ApJ, 799, 96
\bibitem[Gruendl \& Chu (2009)]{Gruendl09}
Gruendl R.\ A., Chu Y.-H., 2009, ApJS, 184, 172
\bibitem[G\"ultekin et al.(2009)]{Gultekin09}
G\"ultekin K., et al., 2009, ApJ, 698, 198
\bibitem[G\"urkan et al.(2015)]{Gurkan15}
G\"urkan G., et al., 2015, arXiv:1507.01552
\bibitem[G\"uver \& \"Ozel (2009)]{Guver09}
G\"uver T., \"Ozel F., 2009, MNRAS, 400, 2050
\bibitem[Hardcastle, Evans \& Croston (2007)]{Hardcastle07}
Hardcastle M.\ J., Evans D.\ A., Croston J.\ H., 2007, MNRAS, 376, 1849
\bibitem[H\"onig et al.(2013)]{Honig13}
H\"onig S.\ F., et al., 2013, ApJ, 771, 87
\bibitem[Hony et al.(2011)]{Hony11}
Hony S., et al., 2011, A\&A, 531, 137 (H11)
\bibitem[Hopkins et al.(2003)]{Hopkins03}
Hopkins A.\ M., et al., 2003, ApJ, 599, 971
\bibitem[Humphrey et al.(2015)]{Humphrey15}
Humphrey A., et al., 2015, MNRAS, 447, 3322
\bibitem[Kaspi et al.(2000)]{Kaspi00}
Kaspi S., Smith P.\ S., Netzer H., Maoz D., Jannuzi B.\ T., Giveon U., 2000,
ApJ, 533, 631
\bibitem[Katkov, Kniazev \& Sil'chenko (2015)]{Katkov15}
Katkov I.\ Yu., Kniazev A.\ Yu., Sil'chenko O.\ K., 2015, AJ, 150, 24
\bibitem[Kauffmann et al.(2003a)]{Kauffmann03a}
Kauffmann G., et al., 2003a, MNRAS, 341, 33
\bibitem[Kauffmann et al.(2003b)]{Kauffmann03b}
Kauffmann G., et al., 2003b, MNRAS, 346, 1055
\bibitem[Kellermann et al.(1989)]{Kellermann89}
Kellermann K.\ I., Sramek R., Schmidt M., Shaffer D.\ B., Green R., 1989, AJ,
98, 1195
\bibitem[Kemper et al.(2010)]{Kemper10}
Kemper F., et al., 2010, PASP, 122, 683
\bibitem[Kewley et al.(2001)]{Kewley01}
Kewley L.\ J., Heisler C.\ A., Dopita M.\ A., Lumsden S., 2001, ApJS, 132, 37
\bibitem[Kobulnicky et al.(2003)]{Kobulnicky03}
Kobulnicky H.\ A., Nordsieck K.\ H., Burgh E.\ B., Smith M.\ P., Percival J.\
W., Williams T.\ B., O'Donoghue D., 2003, SPIE, 4841, 1634
\bibitem[Kormendy \& Ho (2013)]{Kormendy13}
Kormendy J., Ho L.\ C., 2013, ARA\&A, 51, 511
\bibitem[LaMassa et al.(2013a)]{Lamassa13a}
LaMassa S.\ M., Heckman T.\ M., Ptak A., Urry C.\ M., 2013a, ApJ, 765, L33
\bibitem[LaMassa et al.(2013b)]{Lamassa13b}
LaMassa S.\ M., et al., 2013b, MNRAS, 436, 3581
\bibitem[Lyu, Hao \& Li (2014)]{Lyu14}
Lyu J., Hao L., Li A., 2014, ApJ, 792, L9
\bibitem[Marconi \& Hunt (2003)]{Marconi03}
Marconi A., Hunt L.\ K., 2003, ApJ, 589, L21
\bibitem[McConnell \& Ma (2013)]{Mcconnell13}
McConnell N.\ J., Ma C.-P., 2013, ApJ, 764, 184
\bibitem[McDermid et al.(2015)]{Mcdermid15}
McDermid R.\ M., et al., 2015, MNRAS, 448, 3484
\bibitem[McQuillin \& McLaughlin (2013)]{Mcquillin13}
McQuillin R.\ C., McLaughlin D.\ E., 2013, MNRAS, 434, 1332
\bibitem[Meixner et al.(2006)]{Meixner06}
Meixner M., et al., 2006, AJ, 132, 2268
\bibitem[Netzer (2015)]{Netzer15}
Netzer H., 2015, ARA\&A, 53
\bibitem[Oh et al.(2015)]{Oh15}
Oh K., Yi S.\ K., Schawinski K., Koss M., Trakhtenbrot B., Soto K., 2015,
ApJS, 219, 1
\bibitem[Osterbrock (1974)]{Osterbrock74}
Osterbrock D.\ E., 1974, Astrophysics of Gaseous Nebulae, W.H.\ Freeman and
Company
\bibitem[Osterbrock (1977)]{Osterbrock77}
Osterbrock D.\ E., 1977, ApJ, 215, 733
\bibitem[Osterbrock (1981)]{Osterbrock81}
Osterbrock D.\ E., 1981, ApJ, 249, 462
\bibitem[Park et al.(2015)]{Park15}
Park D., Woo J.-H., Bennert V.\ N., Treu T., Auger M.\ W., Malkan M.\ A.,
2015, ApJ, 799, 164
\bibitem[Planck Collaboration (2015)]{Planck15}
Planck Collaboration, 2015, arXiv:1502.01589
\bibitem[Poggianti et al.(2001)]{Poggianti01}
Poggianti B.\ M., et al., 2001, ApJ, 563, 118
\bibitem[Poggianti et al.(2009)]{Poggianti09}
Poggianti B.\ M., et al., 2009, ApJ, 697, L137
\bibitem[Poggianti et al.(2013)]{Poggianti13}
Poggianti B.\ M., et al., 2013, ApJ, 762, 77
\bibitem[Ramos Almeida et al.(2011)]{Ramosalmeida11}
Ramos Almeida C., Dicken D., Tadhunter C., Asensio Ramos A., Inskip K.\ J.,
Hardcastle M.\ J., Mingo B., 2011, MNRAS, 413, 2358
\bibitem[Reines, Greene \& Geha (2013)]{Reines13}
Reines A.\ E., Greene J.\ E., Geha M., 2013, ApJ, 775, 116
\bibitem[Renzini \& Peng (2015)]{Renzini15}
Renzini A., Peng Y., 2015, ApJ, 801, L29
\bibitem[Rose et al.(2013)]{Rose13}
Rose M., Tadhunter C.\ N., Holt J., Rodr\'{\i}guez Zaur\'{\i}n J., 2013,
MNRAS, 432, 2150
\bibitem[Rupke, Veilleux \& Sanders (2005)]{Rupke05}
Rupke D.\ S., Veilleux S., Sanders D.\ B., 2005, ApJ, 632, 751
\bibitem[Salviander \& Shields (2013)]{Salviander13}
Salviander S., Shields G.\ A., 2013, ApJ, 764, 80
\bibitem[Sanghvi et al.(2014)]{Sanghvi14}
Sanghvi J., Kotilainen J.\ K., Falomo R., Decarli R., Karhunen K., Uslenghi
M., 2014, MNRAS, 445, 1261
\bibitem[Savorgnan \& Graham (2015)]{Savorgnan15}
Savorgnan G.\ A.\ D., Graham A.\ W., 2015, MNRAS, 446, 2330
\bibitem[Schartmann et al.(2014)]{Schartmann14}
Schartmann M., Wada K., Prieto M.\ A., Burkert A., Tristram K.\ R.\ W., 2014,
MNRAS, 445, 3878
\bibitem[Scharw\"achter et al.(2015)]{Scharwachter15}
Scharw\"achter J., Combes F., Salom\'e P., Sun M., Krips M., arXiv:1507.02292
\bibitem[Schmidt \& Green (1983)]{Schmidt83}
Schmidt M., Green R.\ F., 1983, ApJ, 269, 352
\bibitem[Spoon et al.(2007)]{Spoon07}
Spoon H.\ W.\ W., Marshall J.\ A., Houck J.\ R., Elitzur M., Hao L., Armus L.,
Brandl B.\ R., Charmandaris V., 2007, ApJ, 654, L49
\bibitem[Stalevski et al.(2012)]{Stalevski12}
Stalevski M., Fritz J., Baes M., Nakos T., Popovi\'c L.\ \v{C}., 2012, MNRAS,
420, 2756
\bibitem[Stasi\'nska et al.(2008)]{Stasinska08}
Stasi\'nska G., Vale Asari N., Cid Fernandes R., Gomes J.\ M., Schlickmann M.,
Mateus A., Schoenell W., Sodr\'e L., 2008, MNRAS, 391, L29
\bibitem[Staveley-Smith et al.(2003)]{Staveleysmith03}
Staveley-Smith L., Kim S., Calabretta M.\ R., Haynes R.\ F., Kesteven M.\ J.,
2003, MNRAS, 339, 87
\bibitem[Tadhunter et al.(1993)]{Tadhunter93}
Tadhunter C.\ N., Morganti R., di Serego-Alighieri S., Fosbury R.\ A.\ E.,
Danziger I.\ J., 1993, MNRAS, 263, 999
\bibitem[Tadhunter et al.(1998)]{Tadhunter98}
Tadhunter C.\ N., Morganti R., Robinson A., Dickson R., Villar-Martin M.,
Fosbury R.\ A.\ E., 1998, MNRAS, 298, 1035
\bibitem[Tasca et al.(2014)]{Tasca14}
Tasca L.\ A.\ M., et al., 2014, A\&A, 564, L12
\bibitem[Tatton et al.(2013)]{Tatton13}
Tatton B.\ L., et al., 2013, A\&A, 554, 33
\bibitem[Trakhtenbrot et al.(2015)]{Trakhtenbrot15}
Trakhtenbrot B., et al., 2015, Science, 349, 168
\bibitem[Tremaine et al.(2002)]{Tremaine02}
Tremaine S., et al., 2002, ApJ, 574, 740
\bibitem[Trujillo et al.(2007)]{Trujillo07}
Trujillo I., Conselice C.\ J., Bundy K., Cooper M.\ C., Eisenhardt P., Ellis
R.\ S, 2007, MNRAS, 382, 109
\bibitem[Urry \& Padovani (1995)]{Urry95}
Urry M., Padovani P., 1995, PASP, 107, 803
\bibitem[van Loon et al.(2013)]{Vanloon13}
van Loon J.\ Th., et al., 2013, A\&A, 550, 108
\bibitem[Vestergaard \& Peterson (2006)]{Vestergaard06}
Vestergaard M., Peterson B.\ M., 2006, ApJ, 641, 689
\bibitem[Voit et al.(2015)]{Voit15}
Voit G.\ M., Bryan G.\ L., O'Shea B.\ W., Donahue M., 2015, arXiv:1505.03592
\bibitem[Volonteri \& Ciotti (2013)]{Volonteri13}
Volonteri M., Ciotti L., 2013, ApJ, 768, 29
\bibitem[Vulcani et al.(2015)]{Vulcani15}
Vulcani B., Poggianti B.\ M., Fritz J., Fasano G., Moretti A., Calvi R.,
Paccagnella A., 2015, ApJ, 798, 52
\bibitem[Wake, van Dokkum \& Franx (2012)]{Wake12}
Wake D.\ A., van Dokkum P.\ G., Franx M., 2012, ApJ, 751, L44
\bibitem[Wijesinghe et al.(2011)]{Wijesinghe11}
Wijesinghe D.\ B., et al., 2011, MNRAS, 410, 2291
\bibitem[Willott, Bergeron \& omont (2015)]{Willott15}
Willott C.\ J., Bergeron J., Omont A., 2015, ApJ, 801, 123
\bibitem[Wong et al.(2015)]{Wong15}
Wong O.\ I., Schawinski K., J\'ozsa G.\ I.\ G., Urry C.\ M., Lintott C.\ J.,
Simmons B.\ D., Kaviraj S., Masters K.\ L., 2015, MNRAS, 447, 3311
\bibitem[Woo et al.(2014)]{Woo14}
Woo J.-H., Kim J.-G., Park D., Bae H.-J., Kim J.-H., Lee S.-E., Kim S.\ C.,
Kwon H.-J., 2014, JKAS, 47, 167
\bibitem[Woods et al.(2011)]{Woods11}
Woods P.M., et al., 2011, MNRAS, 411, 1597
\bibitem[Worthey \& Ottaviani (1997)]{Worthey97}
Worthey G., Ottaviani D.\ L., 1997, ApJS, 111, 377
\bibitem[Wuyts et al.(2012)]{Wuyts12}
Wuyts S., et al., 2012, ApJ, 753, 114
\bibitem[Wyithe \& Loeb (2002)]{Wyithe02}
Wyithe J.\ S.\ B., Loeb J., 2002, ApJ, 581, 886
\bibitem[Yang, Chen \& Huang (2015)]{Yang15}
Yang X.-H., Chen P.-S., Huang Y., 2015, MNRAS, 449, 3191
\end{thebibliography}
\end{document}